\documentclass[prd,aps,preprint,tightenlines,showpacs,superscriptaddress,nofootinbib]{revtex4-1}

\usepackage{epsfig}
\usepackage{amsmath}

\begin{document}

\title{Sudakov Resummation in Small-$x$ Saturation Formalism}

\author{A. H. Mueller}
\affiliation{Department of Physics, Columbia University, New York, NY 10027, USA}

\author{Bo-Wen Xiao}
\affiliation{Institute of Particle Physics, Central China Normal University, Wuhan 430079, China}
\affiliation{Department of Physics, Pennsylvania State University, University Park, PA 16802, USA}

\author{Feng Yuan}
\affiliation{Nuclear Science Division, Lawrence Berkeley National
Laboratory, Berkeley, CA 94720, USA}

\begin{abstract}
Through an explicit calculation of massive
scalar particle (e.g., Higgs boson) production in high energy $pA$ collisions up to
one-loop order, we demonstrate that the Sudakov-type logarithms in hard processes
in small-$x$ saturation formalism can be systematically separated from the typical small-$x$ logarithms. The generic feature of the Sudakov logarithms and all order 
resummation is derived. We further comment on the phenomenological implications and
extension to other hard processes in the small-$x$ calculations. 
\end{abstract}
\maketitle


{\it 1. Introduction.}
An important application of the perturbative quantum 
chromodynamics (QCD) is the resummation.
In high energy hadronic processes involving large separate
scales, resummation is not only necessary to make
reliable predictions, but also crucial to extract the fundamental
properties of the strong interaction theory, such as the 
strong coupling constant. In practice, the resummation formalism
has been applied to a wide range of physics processes.
One of the examples is the resummation of the Sudakov-type
double logarithms~\cite{Sudakov:1954sw,Collins:1984kg}. 
The double logarithms appear in, for example, the transverse momentum
spectrum of a hard process, where each order
of perturbative correction is accompanied by a large double logarithmic
term of $\ln^2( Q^2/k_\perp^2)$ with $Q$ the large
momentum scale and $k_\perp$ the transverse momentum.
In low transverse momentum region where most of the
production events sit,  the QCD resummation has to 
be performed. This resummation is often
referred as the transverse momentum resummation.

Meanwhile, there is also small-$x$ resummation which 
is equally important, in particular,
in the area of the large hadron collider (LHC). 
Small-$x$ resummation is governed by the 
BFKL evolution~\cite{Balitsky:1978ic}. 
Because of high gluon density in nucleon/nucleus, 
the non-linear term in the evolution plays a very important role at 
small-$x$~\cite{Gribov:1984tu,Mueller:1985wy,McLerran:1993ni}, 
which leads to the BK-JIMWLK evolution~\cite{Balitsky:1995ub,JalilianMarian:1997jx,Iancu:2000hn}.
As a result, the gluon saturation with a characteristic scale $Q_s^2$
becomes inevitable at small-$x$. Seeking for the signal of
the gluon saturation phenomena and studying the
associated dynamics has been one of the most
important motives for the high energy nucleon-nucleus
experiments at RHIC and the LHC, and for the 
planed electron-ion colliders. Great efforts have been
made from both experiment and theory sides~\cite{eic}.
Among them, the hard processes involving a large
momentum scale have been emphasized recently as
important probes for the saturation phenomena~\cite{Dominguez:2010xd}. 
This is because these hard processes can directly measure the transverse
momentum dependence of the gluon distributions,
whose behavior manifest the saturation phenomena~\cite{Mueller:1985wy,McLerran:1993ni}.
However, these hard processes also impose a question: are there
Sudakov double logarithms as well? If yes, how to resum these
double logarithms consistently in the small-$x$ saturation
formalism?  To our knowledge, there has been no 
theoretical  investigation on this topic. 

In this paper, we will study, for the first time, the Sudakov-type 
double logarithms in the small-$x$ formalism. 
As an important first step, we take the massive (with mass $Q$) color neutral
scalar particle production in $pA$ collisions as an example. 
The scalar particle is directly coupled to the gluon 
through an effective lagrangian ${\cal L}_{eff}=-\frac{1}{4}g_\phi\Phi F^{a}_{\mu\nu}F^{a\mu\nu} $,
where $\Phi$ represents the scalar field, $F^{a\mu\nu}$ is the field
tensor for the gluon field with the associated color index $a$.
This effective lagrangian has been used to calculate the Standard Model
Higgs boson production in hadronic collisions~\cite{Dawson:1990zj}, and to study the
gluon saturation in nucleus~\cite{CU-TP-441a} as well.
We extend the previous calculations to the scalar particle production in
$pA$ collisions in the small-$x$ factorization formalism up to
one-loop order. At this order, we will be able to identify the 
Sudakov double logarithms, which is absent in the leading order
evaluation. As a result, all order resummation can
be performed consistently with the small-$x$ evolution.
One of the important steps to achieve this is to separate the soft gluon
radiation (which contributes to the Sudakov logs) from those
contributing to the small-$x$ evolution. Our explicit
one-loop calculations will demonstrate that we can consistently resum
Sudakov double logarithms and the BFKL (BK-JIMWLK) evolution
at the same time.
This will provide important guidelines for 
further developments in other hard processes 
which are crucial to study the saturation 
phenomena in the small-$x$ physics. 

The rest of the paper is organized as follows. In Sec.II, We will
present the calculation of scalar particle 
production in $pA$ collisions in the small-$x$ 
formalism up to one-loop order. The Sudakov double
logarithms are identified. All order resummation 
is derived in Sec.III. We summarize our results and present
further discussions in Sec.IV.

{\it 2. Massive Scalar Production in $pA$ Collisions at One-loop Order.}
We follow the high energy small-$x$ factorization formalism to 
calculate the massive scalar particle production in $pA$ collisions which implies 
that the center of mass energy $S \gg Q^2$ while $Q^2 \gg k_\perp^2$.
Similar to those calculated in Ref.~\cite{Chirilli:2011km}, the gluon distribution from 
the nucleon is taken to be collinear, therefore
the transverse-momentum distribution of the produced
scalar particle reflects the transverse-momentum dependence of the
gluon distribution in the nucleus. Due to the high gluon
density inside the target nucleus, the multiple interactions must be taken into account.

The leading order contribution can be easily formulated
following the small-$x$ factorization formalism~\cite{CU-TP-441a},
\begin{equation}
\frac{d\sigma^{(\rm LO)}}{dyd^2k_\perp}=\sigma_0\int \frac{d^2x_\perp d^2x_\perp'}{(2\pi)^2}e^{ik_\perp\cdot r_\perp}
xg_p(x)S^{WW}_Y(x_\perp,x_\perp') \ ,
\end{equation}
where $r_\perp=x_\perp-x_\perp'$, $\sigma_0=g_\phi^2/g^232(1-\epsilon)$ with $\epsilon=-(D-4)/2$
the dimensional regulation parameter, 
$y$ and $k_\perp$ are the rapidity and transverse momentum
for the scalar particle, $g_p(x)$ denotes the integrated gluon distribution
from the nucleon, and $S^{WW}$ represents the unintegrated gluon
distribution function from the nucleus. Here, $x\approx Qe^y/\sqrt{S}$ and $x_a\approx
Qe^{-y}/\sqrt{S}$ are the momentum fractions of incoming gluons from the proton projectile and target nucleus, respectively. 
Because of the colorless nature of the scalar particle, there are only the initial
state interactions for the multiple gluon exchange with the nucleus target.
Therefore, the associated un-integrated gluon distribution is the so-called
Weizs\"{a}cker-Williams (WW) gluon distribution in the small-$x$ formalism~\cite{Dominguez:2010xd},
\begin{equation}
S^{WW}_Y(x_\perp,y_\perp)=-\left\langle{\rm Tr}\left[ \partial_\perp^\beta U(x_\perp)
U^\dagger(y_\perp)\partial_\perp^\beta U(y_\perp)U^\dagger(x_\perp)\right]\right\rangle_Y \ ,
\end{equation}
where $Y$ represents the rapidity of the gluon from the nucleus $Y\sim \ln (1/x_a)$,
and sum over the transverse index $\beta$ is implicit. 
The so-called linearly polarized gluon distribution can also be taken into
account where different projection of the transverse index is performed~\cite{Metz:2011wb}. 
The Wilson line $U$ is defined as 
$U(x_\perp)={\cal P}\exp\left(-ig\int_{-\infty}^{+\infty} dx^-A^+(x^-,x_\perp)\right)$.

\begin{figure}[tbp]
\begin{center}
\includegraphics[width=6.3cm]{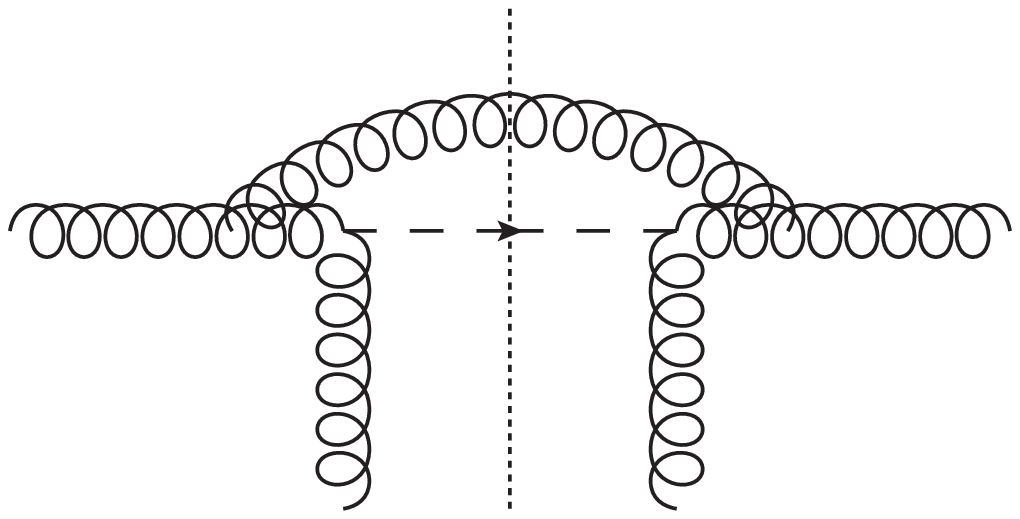}
\includegraphics[width=7cm]{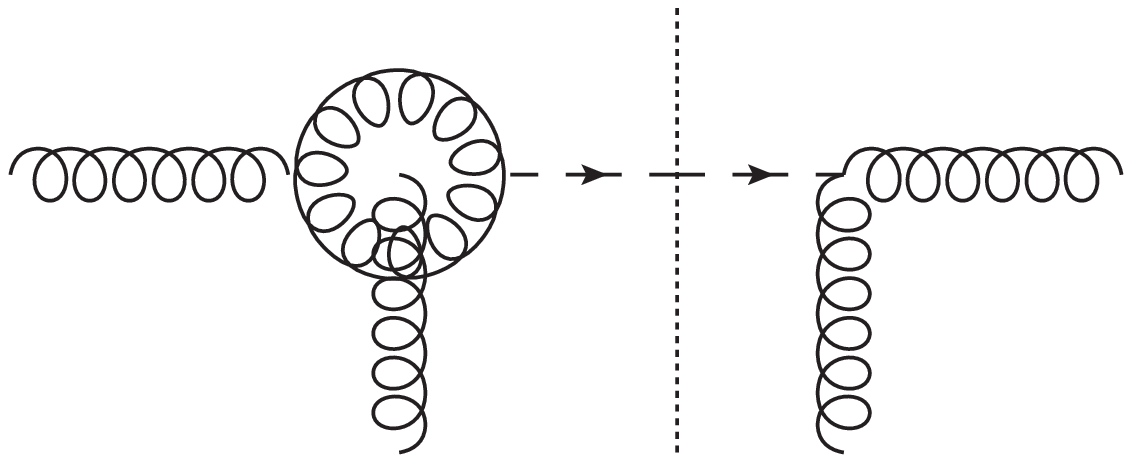}
\end{center}
\caption[*]{Examples of one-loop diagrams from the real gluon radiation (L)
and virtual gluon radiation (R) for the scalar particle production 
in $pA$ collisions in the saturation formalism with the multiple interactions with 
the nucleus target taken into account. Here the vertical gluon represents arbitrary number of the WW small-$x$ gluons which are summed into Wilson lines.}
\label{real}
\end{figure}

At one-loop order,  real and virtual diagrams yield higher order corrections which contain various divergences. In addition to the rapidity divergence associated with the WW gluon distribution
from the nucleus, there is also the collinear divergence for the integrated gluon
distribution from the nucleon. In our calculations, the dimensional regulation
($D=4-2\epsilon$) is used to formulate the collinear divergence, whereas the direct subtraction is applied for the rapidity divergence. In Fig.~\ref{real}, we plot the typical diagrams from the
real and virtual gluon radiations at one-loop order . 
The scattering amplitude from the real diagrams can be written as,
\begin{eqnarray}
{\cal A}^{\nu\alpha}&=&\int\frac{d^2k_{g1\perp}}{(2\pi)^2}\frac{zk_{1\perp}^\alpha g^{\nu\beta}+(1-z)k_{1\perp}^\beta g^{\alpha\nu}
-z(1-z)k_{1\perp}^\nu g^{\alpha\beta}}{k_{1\perp}^2}\Gamma_A^\beta(k_{g1\perp},k_{g2\perp})\nonumber\\
&&+\left[\left(\frac{zk_\perp^{\prime\alpha}k_\perp^{\prime\nu}}{k_\perp^{\prime 2}+\epsilon_f^2}-
\frac{zk_\perp^{\alpha}k_\perp^{\nu}}{k_\perp^{2}+\epsilon_f^2}\right)
+z(1-z)Q^2\frac{g^{\alpha\nu}}{2}\left(\frac{1}{k_\perp^{2}+\epsilon_f^2}-\frac{1}{k_\perp^{\prime 2}
+\epsilon_f^2}\right)\right]\Gamma_B(k_{g\perp})\ ,
\end{eqnarray}
where $k_\perp$ and $k_{2\perp}$ represent the transverse momenta for the final state 
scalar particle and the radiated gluon, respectively,
$k_{g\perp}=k_{g1\perp}+k_{g2\perp}=k_\perp+k_{2\perp}$, 
$\epsilon_f^2=(1-z)Q^2$ with $z$ the momentum fraction of the incoming gluon
carried by the scalar particle. For convenience, we further define 
$k_{1\perp}=k_\perp-k_{g1\perp}$, $k_\perp'=k_\perp-zk_{g\perp}$,
and $\nu$ and $\alpha$ are the polarization vector indices for the incoming 
and outgoing gluons, for which we have chosen the physical polarizations. 
$\Gamma_A$ and $\Gamma_B$ are defined as
\begin{eqnarray}
\Gamma_A^\beta(k_{g1\perp},k_{g2\perp})&=&\int d^2x_1d^2x_2 e^{ik_{g1\perp}\cdot x_1+ik_{g2\perp}\cdot x_2}
{\rm Tr}\left[U^\dagger(x_2)T^bU(x_2)[i\partial_\perp^\beta U^\dagger(x_1)U(x_1),T^a]\right] \ , \\
\Gamma_B(k_{g\perp})&=&2\int d^2x_\perp e^{ik_{g\perp}\cdot x_\perp}{\rm Tr}\left[T^bU(x_\perp)T^aU^\dagger(x_\perp)\right] \ ,
\end{eqnarray}
where $a$ and $b$ are color indices for the incoming and outgoing gluons, respectively. 
Clearly, the amplitude squared from the above expressions will depend
on multi-gluon correlation functions (beyond the WW-gluon distribution) 
from the nucleus, as this is the common feature in the high order
calculations in the small-$x$ formalism~\cite{Metz:2011wb}. 
However, in the $k_\perp\ll Q$ limit, these correlation
functions is either reduced to the WW-gluon distribution, or 
absorbed into the evolution of the WW-gluon distribution. 
To evaluate the contribution from the real gluon radiation, 
we integrate out the phase space of the radiated gluon ($k_2$). 
To simplify the calculation, we perform the power expansion of the 
amplitude squared in terms of $k_\perp/Q$, and only keep
the leading power contributions.
In the power counting analysis, we find that 
the phase space integral contains three important 
contributions: (1) soft gluon radiation $k_2^+\sim k_2^-\sim k_{2\perp}$ 
which eventually leads to the Sudakov logarithms; (2) collinear gluon
contribution as respect to the incoming nucleon projectile;
(3) collinear gluon contribution as respect to the target nucleus.
The soft gluon contribution can be easily obtained 
in the limit of $Q^2\gg k_\perp^2$, which results into 
$\delta(k_2^+)\delta(k_2^-)\ln({Q^2}/{k_{2\perp}^{2}})$.
When Fourier transformed into the impact parameter space,
this term leads to a soft divergence in terms of $1/\epsilon^2$ in
dimension regulation. The soft divergence will be cancelled by the
relevant virtual diagrams. The last contribution contains the rapidity divergence and gives rise to the evolution of the WW gluon distribution.

The evaluation of the virtual diagrams leads to the following contributions,
\begin{eqnarray}
&&-\alpha_s\int \frac{dz}{z(1-z)}\frac{d^2q_\perp d^2k_{g1\perp}}{(2\pi)^4}\Gamma_C(k_{g1\perp},k_{g2\perp})
\left[\frac{k_{g1\perp}^\nu}{q_\perp^2}-\frac{2q_\perp'^\nu q_\perp\cdot q_\perp'-q_\perp^{\nu}q_\perp^{\prime 2}}
{q_\perp^2 (q_\perp^{\prime 2}+\epsilon_f^{\prime 2})} \right]\nonumber\\
&&+\frac{\alpha_sN_c}{\pi}\beta_0\left(\frac{1}{\epsilon_{UV}}-\frac{1}{\epsilon_{IR}}\right)\Gamma_D(k_\perp)\  ,\label{virtual}
\end{eqnarray}
where $\epsilon_f^{\prime 2}=-z(1-z)Q^2$, $q_\perp'=q_\perp+k_{g1\perp}-zk_{\perp}$ with $k_{\perp}=k_{g1\perp}+k_{g2\perp}$,
$\beta_0=11/12-N_f/18$ with $N_f$ number of flavors, 
$\epsilon_{UV}$ and $\epsilon_{IR}$ represent the ultra-violet and
infrared divergences in the loop diagrams, respectively, and 
$\Gamma_C$ and $\Gamma_D$ are defined as 
\begin{eqnarray}
\Gamma_C(k_{g1\perp},k_{g2\perp})&=&
\int d^2 x_1 d^2x_2 e^{ik_{g1\perp}\cdot x_1+ik_{g2\perp}\cdot x_2}\nonumber\\
&&\times\left\{{\rm Tr}\left[U(x_1)T^aU^\dagger(x_2)\right]{\rm Tr}\left[U^\dagger(x_1)U(x_2)\right]-(x_1\leftrightarrow x_2)\right\} \ ,\\
\Gamma_D^\beta(k_\perp)&=&\int d^2 x e^{ik_{g\perp}\cdot x_\perp}
{\rm Tr}\left[T_ai\partial_\perp^\beta U^\dagger(x_\perp)U(x_\perp)\right] \ .
\end{eqnarray}
In Eq.~(\ref{virtual}), the ultra-violet (UV) divergences cancel out between the two contributions in the first term, whereas the remaining UV-divergence in the second term is normally interpreted as the charge renormalization, which appears in the form of
$({\alpha_s}/{\pi})N_c\beta_0\left(-{1}/{\epsilon_{UV}}+\ln({Q^2}/{\mu^2})\right)$.
Furthermore, there is also IR divergence, which can be
absorbed into the renormalization for the incoming gluon distribution from the
nucleon. In addition, there are rapidity divergence and soft 
divergence in the first term of Eq.~(\ref{virtual}). The soft divergence is regulated by
dimension regularization, and is found to proportional to $1/\epsilon^2$. This divergence,
as mentioned above, will cancel out that from the real diagrams. 

After canceling out the soft divergences from the real and virtual diagrams, we are left with the 
rapidity divergence and collinear divergence,
\begin{eqnarray}
xg_p(x)\int\frac{d\xi}{\xi}{\bf K}_{\rm DMMX}\otimes S^{WW}(x_\perp,y_\perp)+\left(-\frac{1}{\epsilon}
\right)\frac{\alpha_s N_c}{\pi}S^{WW}(x_\perp,y_\perp){\cal P}_{g/g}\otimes xg_p(x) \ ,
\end{eqnarray}
where ${\bf K}_{\rm DMMX}$ is the BK-type of evolution kernel for the WW
gluon distribution, and ${\cal P}_{g/g}=\frac{1}{(1-z)_+}+\frac{1}{z}+z(1-z)+\beta_0\delta(1-z)$ the collinear evolution
kernel for the incoming gluon from the nucleon. 
Our result for ${\bf K}_{\rm DMMX}$ is consistent with a recent
calculation for the evolution directly from the JIMWLK
evolution~\cite{Dominguez:2011gc}. After subtracting these divergences by the
renormalizations of the associated distributions, 
we obtain the final result for the differential
cross section as
\begin{eqnarray}
\frac{d\sigma^{\rm (LO+NLO)}}{dyd^2k_\perp}|_{k_\perp\ll Q}&=&\sigma_0
\int \frac{d^2x_\perp d^2x_\perp'}{(2\pi)^2}e^{ik_\perp\cdot r_\perp}
S^{WW}_{Y=\ln 1/x_a}(x_\perp,x_\perp')xg_p(x,\mu^2=\frac{c_0^2}{r_\perp^2})\nonumber\\
&&\left\{1+\frac{\alpha_s}{\pi} N_c \left[\beta_0\ln\frac{Q^2r_\perp^2}{c_0^2}-\frac{1}{2}\left(\ln\frac{Q^2r_\perp^2}{c_0^2}\right)^2
+\frac{\pi^2}{2}\right]\right\}\ ,
\end{eqnarray}
where $c_0=2e^{-\gamma_E}$ with $\gamma_E$ the Euler constant.
We have also introduced the scale for the integrated gluon distribution 
$\mu^2=c_0^2/r_\perp^2$ and the rapidity for 
the WW gluon $Y=\ln 1/x_a$ to simplify the above expression.

Comparing to the differential cross section calculation for inclusive
hadron production in $pA$ collisions~\cite{Chirilli:2011km}, we find 
that they share the same structure. In particular, the integrated parton
distribution is more convenient to be set at the scale of $c_0/r_\perp$. 
The major difference is that, the above formula is valid in the limit of $Q\gg k_\perp$
where we have to resum the Sudakov logarithms. That is also the reason
that the un-integrated gluon distribution from the nucleus side does not 
depend on more complicated structure of the Wilson lines. 
If we calculate the differential cross section in the kinematic region
of $k_\perp\sim Q$, the multi-gluon correlation functions (beyond the
WW-gluon distribution) from the nucleus will contribute, similar to what 
was found in Ref.~\cite{Chirilli:2011km}. 
Those contributions, however, is power suppressed in the limit of
$k_\perp\ll Q$.

We have also performed this calculation in coordinate 
space with cutoffs putting into integrals instead of the use of the dimensional 
regularization, and obtained the same results. This implies that there is no 
scheme dependence in the Sudakov logarthms, at least to the level we are concerned.

{\it 3. All Order Resummation.}
Our calculations at one-loop
order in the above demonstrate that the soft gluon radiation is well separated
from the collinear gluon radiations. In particular, the soft gluon radiation comes
from the initial state radiation, and in the $Q\gg k_\perp$ limit, the relevant gluon
distribution from the nucleus takes the form of the WW gluon distribution. Higher order
soft gluon radiations will follow the same form. 
This can be understood that the Sudakov double logarithms come from 
soft gluon radiation associated with the hard probe. To resum these large logarithms, we 
follow the Collins-Soper-Sterman procedure~\cite{Collins:1984kg}. In particular, we can 
write down an evolution equation respect to the hard scale $Q^2$. 
By solving the differential equation, we can resum the differential 
cross section in terms,
\begin{eqnarray}
\frac{d\sigma^{\rm ({\rm resum})}}{dyd^2k_\perp}|_{k_\perp\ll Q}&=&\sigma_0
\int \frac{d^2x_\perp d^2x_\perp'}{(2\pi)^2}e^{ik_\perp\cdot r_\perp}e^{-{\cal S}_{sud}(Q^2,r_\perp^2)}
S^{WW}_{Y=\ln 1/x_a}(x_\perp,x_\perp')\nonumber\\
&&\times xg_p(x,\mu^2={c_0^2}/{r_\perp^2})\left[1+\frac{\alpha_s}{\pi}\frac{\pi^2}{2}N_c\right]\ ,\label{resum}
\end{eqnarray}
where the Sudakov form factor contains all order resummation 
\begin{eqnarray}
{\cal S}_{sud}=\int_{C_1^2/r^2}^{C_2^2Q^2}\frac{d\mu^2}{\mu^2}\left[A\ln\frac{C_2^2Q^2}{\mu^2}+B\right] \ , 
\end{eqnarray}
where $C_{1,2}$ are parameters in order of 1.
The hard coefficients $A$ and $B$ can be
calculated perturbatively: $A=\sum\limits_{i=1}^\infty A^{(i)}\left(\frac{\alpha_s}{\pi}\right)^i$. 
From the explicit results for the one-loop calculations, we find that they are
\begin{equation}
A^{(1)}=N_c, ~~~B^{(1)}=-\beta_0 N_c\ ,
\end{equation}
where we have chosen the so-called canonical variables 
for $C_1=c_0$ and $C_2=1$. The above formulas are the final 
results for the soft gluon resummation in massive scalar particle production
in $pA$ collisions. The result is valid in the limit of $k_\perp\ll Q$, 
and we have applied the small-$x$ factorization where higher order
in $1/\ln(1/x_a)$ has been neglected as well. As compared to Ref.~\cite{Ji:2005nu} which takes the $\alpha_s$ correction of $g_\phi$ into account, we find the above coefficients are consistent with the known results\cite{Berger:2002ut} except for $B^{(1)}$ which misses a factor of 2. This difference is due to the convention in the saturation formalism in which we do not include the virtual gluon and quark loops contribution for the vertical WW small-$x$ gluons as shown in Fig.~\ref{real}. 

We would like to emphasize a number of important features in 
the above derivation. First, the collinear divergence
and the soft divergence is well separated. This follows the transverse
momentum resummation in the collinear factorization. 
Second and most importantly, the rapidity divergence from the
the un-integrated gluon distribution of the nucleus is also well separated
from the soft gluon radiation. This is because the rapidity divergence
comes form the collinear gluon radiation parallel to the nuclei momentum,
whereas the Sudakov logarithms come from soft gluon region. 

Third, as the above features are common features
in the small-$x$ saturation formalism, we expect our resummation
results can be extended to other hard processes. An immediate
application is the color neutral particle production in $pA$ 
collisions, such as the two-photon production~\cite{Qiu:2011ai} and heavy quarkonium
production~\cite{Sun:2012vc}. There are only initial state interactions 
in these processes, and the Sudakov resummation will be the same as
what we have derived in this paper for the massive scalar particle
production. We expect the similar resummation formula.
Another important extension is the dijet correlation
in $pA$ collisions~\cite{Dominguez:2010xd}, which has attracted great attentions in recent
years. Of course, because the final state of dijet production carries 
color, we have to take into account the final state interaction
effects in the Sudakov resummation. Therefore, the resummation 
formula will be different from the scalar particle production studied
in this paper. However, the generic form will remain the same.
We plan to address this calculation in a future publication.

{\it 4. Discussions and Conclusion.}
In summary, we have demonstrated the Sudakov double
logarithms in the small-$x$ calculations of hard processes
in $pA$ collisions. All order resummation formalism has
been derived. This result will have potential application
for other hard processes in $pA$ collisions, from which
we hope to investigate the saturation physics.

In addition, the technique developed in this paper shall
also be relevant for the hard processes in hot/dense medium,
such as the quark-gluon plasma. In particular, in the hard 
processes of jet penetrating through the medium, we expect
the similar formalism for the Sudakov resummation
effects. The direct consequence is that the transverse 
momentum broadening in the hard processes (with
hard momentum scale much larger than the momentum
scale in the medium) will be dominated by the Sudakov logarithms
which is the same as that in the vacuum. This may lead
to a natural explanation for the azimuthal angle 
correlation of dijet production in heavy ion collisions
which was found the same as that in $pp$ collisions 
in the large angle region~\cite{Aad:2010bu}, where Sudakov effects dominate.

This work was supported in part by the U.S. Department of Energy under the 
contracts DE-AC02-05CH11231 and DOE OJI grant No. DE - SC0002145. We thank J. W. Qiu for comments and discussions.

\end{document}